\documentclass[twocolumn,showpacs,preprintnumbers,amsmath,amssymb]{revtex4}

\usepackage{graphicx}
\usepackage{dcolumn}
\usepackage{bm}

\begin{document}

\title{Topological Aspect of Knotted Vortex Filaments in Excitable Media}\thanks{
Supported by the National Natural Science Foundation of China and
Cuiying programme of Lanzhou University.}

\author{Ji-Rong Ren}
\author{Tao Zhu}\thanks{Corresponding author}\email{zhut05@lzu.cn}
\author{Yi-Shi Duan }

\affiliation{Institute of Theoretical Physics, Lanzhou University,
Lanzhou 730000, China}

\begin{abstract}
Scroll waves exist ubiquitously in three-dimensional excitable
media. It's rotation center can be regarded as a topological object
called vortex filament. In three-dimensional space, the vortex
filaments usually form closed loops, and even linked and knotted. In
this letter, we give a rigorous topological description of knotted
vortex filaments. By using the $\phi$-mapping topological current
theory, we rewrite the topological current form of the charge
density of vortex filaments and use this topological current we
reveal that the Hopf invariant of vortex filaments is just the sum
of the linking and self-linking numbers of the knotted vortex
filaments. We think that the precise expression of the Hopf
invariant may imply a new topological constraint on knotted vortex
filaments.
\end{abstract}

\pacs{02. 10. Kn, 82. 40. CK, 47. 54. -r}

\maketitle

Scroll waves exist ubiquitously in three-dimensional excitable
media. They have been directly observed in three-dimensional
chemical\cite{chem}, physical\cite{phys-CO,phys}, and biological
systems\cite{bio}. Specific examples include the
Belousov-Zhabotinskii reaction\cite{BZ}, the cardiac muscle\cite{car
mus}, and the oxidation of CO on platinum\cite{phys-CO}. Recently,
scroll waves have drawn great interest and have been studied
intensively in many ways because it is believed to be the mechanism
of some re-entrant cardiac arrhythmias and fibrillation which is the
leading cause of death in the industrialized world\cite{7-1,7,8,9}.
The investigation of properties of scroll waves in excitable media
provides insights into the possible behaviors of these processes in
the cardiac tissue of animals whose heart wall is thick enough for
three-dimensional effects to be significant.

On the other hand, the global behavior of a scroll wave is well
described by the motion of its approximate rotation center which is
a line defect known as a vortex filament and usually defined in
terms of a phase singularity. In three-dimensional excitable media,
the vortex filaments is commonly a closed ring, and these vortex
filaments can form linked and knotted rings which contract to
compact, particle-like
bundles\cite{knot1,knot2,knot4,knot5,knot6,topo2,topo3,topo4,knot3,topo1,topo5}.
The existence of the knotted vortex filaments can be regarded as a
topological phenomena in excitable media. In most previous
works\cite{topo2,topo3,topo4,knot3,topo1,topo5}, the topological
arguments have been applied to help us understand the topological
properties of the vortex filaments and some most important
topological constraints on behaviors of the vortex filaments have
been investigated. These topological rules may have some important
applications in practice. In particular, the topological constraint
on knotted vortex filaments is believed to relate to topological
characteristic numbers of knotted vortex filament family, such as
the winding, the self-linking and the linking
numbers\cite{knot3,topo1,topo5}. So in research into knotted vortex
filaments in excitable media, one should play much attention to
these knot characteristics. In this letter, we will use the
topological viewpoint to study the knotted vortex filaments with the
Hopf invariant\cite{hopf} which can be used to describe the linkages
of knot family in mathematics, and reveal the inner relationship
between the Hopf invariant and the topological characteristic
numbers of the knotted vortex filaments. We think that this inner
relationship may imply a new topological constraint on knotted
vortex filaments.

We chose to work with a general two-variable reaction-diffusion
system which mathematical description in terms of a nonlinear
partial differential equation. This equation is written as
\begin{equation}
\partial_t u=f(u,v)+D_u \nabla^2 u,~~\partial_t v=g(u,v)+D_v
\nabla^2v,
\end{equation}
where $u$ and $v$ represent the concentrations of the reagents;
$\nabla^2$ is the Laplacian operator in three-dimensional space;
$f(u,v)$ and $g(u,v)$ are the reaction functions. Following the
description in Ref.\cite{topo2}, we define a complex function
$Z=\phi^1+i\phi^2$, where $\phi^1=u-u^*$ and $\phi^2=v-v^*$. Here
$u^*$ and $v^*$ are the concentrations of the vortex filaments.

As pointed out in Ref.\cite{zero} , the sites of the vortex
filaments are just the isolated zero lines of the complex function
$Z=\phi^1+i\phi^2$ in three-dimensional space. By using the
topological viewpoints, Zhang et al.\cite{topo2} derived a
topological expression of the charge density of these zero lines,
i.e., the vortex filaments, which is written as
\begin{equation}
\vec{\rho}(\vec{x},t)=\sum_{l=1}^{N} W_l \int_{L_l} d\vec{x}
\delta^3(\vec{x}-\vec{x}_l),\label{2}
\end{equation}
where $W_l$ is the topological charge of the $l$-th vortex filament
$L_l$.This expression reveals the topological structures of vortex
filaments. In our topological theory of knotted vortex filaments,
the topological structure of the vortex filaments will play an
essential role. In order to study the knotted vortex filaments more
conveniently, we firstly rewrites the charge density of vortex
filaments as a topological current. Now we begin to derive the
topological current form of the charge density of vortex filaments.
We know that the complex function $Z=\phi^1+i\phi^2$ can be regarded
as the complex representation of a two-dimensional vector field
$\vec{Z}=(\phi^1,\phi^2)$. Let us define the unit vector:
$n^a=\frac{\phi^a}{\|\phi\|} (a=1,2; \|\phi\|^2=\phi^a\phi^a=Z^*Z)$.
It is easy to see that the zeros of $Z$ are just the singularities
of $\vec{n}$. Using this unit vector $\vec{n}$, we define a
topological current
\begin{equation}
j^i=\frac{1}{2\pi}\epsilon^{ijk}\epsilon_{ab}\partial_j n^a
\partial_k n^b,~~~~i,j,k=1,2,3.
\end{equation}
Apply the $\phi$-mapping theory\cite{optic,topo-cur1,topo-cur2}, one
can obtain
\begin{equation}
j^i=\delta(\vec{\phi})D^i(\frac{\phi}{x}),
\end{equation}
where
$D^i(\frac{\phi}{x})=\frac{1}{2}\epsilon^{ijk}\epsilon_{ab}\partial_j
\phi^a \partial_k \phi^b$ is the Jacobian vector. This delta
function expression of the topological current $j^i$ tells us it
doesn't vanish only when the vortex filaments exist, i.e., $Z=0$.
The sites of the vortex filaments determine the nonzero solutions of
$j^i$. The implicit function theory shows that under the regular
condition\cite{imp}
\begin{equation}
D^i(\frac{\phi}{x})\neq 0,
\end{equation}
the general solutions of
\begin{equation}
\phi^1(\vec{x},t)=0,~~\phi^2(\vec{x},t)=0
\end{equation}
can be expressed as
\begin{equation}
x^1=x^1_l(t,s),~~x^2=x^2_l(t,s),~~x^3=x^3_l(t,s),~~l=1,2,\cdot,N,
\end{equation}
which represent the world surface of $N$ moving isolated vortex
filaments $L_l(l=1,2,\cdot\cdot\cdot,N)$ with string parameter $s$.
In delta function theory\cite{del}, one can prove that in
three-dimensional space,
\begin{equation}
\delta(\vec{\phi})=\sum_{l=1}^{N} \beta_k \int_{L_l}
\frac{\delta^3(\vec{x}-\vec{x}_l(s))}{|D(\frac{\phi}{u})|_{\Sigma_l}}ds,\label{8}
\end{equation}
where
$D(\frac{\phi}{u})=\frac{1}{2}\epsilon^{jk}\epsilon_{mn}\frac{\partial
\phi^m}{\partial u^j} \frac{\partial \phi^n}{\partial u^k}$ and
$\Sigma_l$ is the $l$-th planar element transverse to $L_l$ with
local coordinates $(u^1,u^2)$. The positive integer $\beta_l$ is the
Hopf index of $\phi$-mapping, which means that when $\vec{x}$ covers
the neighborhood of the zero point $\vec{x}_l(s,t)$ once, the vector
field $\vec{\phi}$ covers the corresponding region in $\phi$ space
for $\beta_l$ times. Meanwhile the direction vector of $L_l$ is
given by\cite{topo-cur1,topo-cur2}
\begin{equation}
\frac{dx^i}{ds}|_{\vec{x}_l}=\frac{D^i(\phi/x)}{D(\phi/u)}|_{\vec{x}_l}.\label{9}
\end{equation}
Then considering Eqs.(\ref{8}) and Eqs.(\ref{9}), we obtain the
inner structure of $j^i$,
\begin{eqnarray}
j^i&=&\delta(\vec{\phi})D^i(\frac{\phi}{x})\nonumber\\
&=&\sum_{l=1}^{N} \beta_l \eta_l \int_{L_l} dx^i
\delta^3(\vec{x}-\vec{x}_l),\label{10}
\end{eqnarray}
where $\eta_l=sgn D(\frac{\phi}{u})=\pm 1$ is the Brouwer degree of
$\phi$-mapping, with $\eta_l=1$ corresponding to the vortex filament
and $\eta_l=-1$ corresponding to the antivortex filament. Compare
Eqs.(\ref{2}) and Eqs.(\ref{10}) we see that the topological current
$\vec{j}$ is just the charge density vector $\vec{\rho}$ of the
vortex filament in Ref.\cite{topo2}. In our theory, the topological
charge of the vortex filament $L_l$ is
\begin{equation}
Q_l=\int_{\Sigma_l} \vec{j}\cdot d\vec{\sigma}=W_l=\beta_l
\eta_l,\label{11}
\end{equation}
in which $W_l$ is just the winding number of $\vec{Z}$ around $L_l$,
the above expression reveals distinctly that the topological charge
of vortex filament is not only the winding number, but also
expressed by the Hopf indices and Brouwer degrees. The topological
inner structure showed in Eq.(\ref{11}) is more essential than in
Eq.(\ref{2}), this is just the advantage of our topological
description of the vortex filaments.

Now let us begin to discuss the topological properties of knotted
vortex filaments in excitable media. It is well know that the Hopf
invariant is an important topological invariant to describe the
topological characteristics of the knot family. In our topological
theory of knotted vortex filaments, the Hopf invariant relates to
the topological characteristics numbers of the knotted vortex
filaments family. In a closed three-manifold $M$ the Hopf invariant
is defined as\cite{optic,hopf}
\begin{equation}
H=\frac{1}{2\pi}\int_{M}A_ij^id^3x,\label{12}
\end{equation}
in which $A_i$ is a "induced Abelian gauge potential" constructed
with the complex function $Z$. The relationship between the Abelian
gauge field strength $F_{ij}=\partial_i A_j-\partial_j A_i$ and the
topological current $j^i$ is\cite{topo-cur2}
\begin{equation}
j^i=\frac{1}{4\pi}\epsilon^{ijk}F_{jk}=\frac{1}{2\pi}\epsilon^{ijk}\epsilon_{ab}\partial_j
n^a\partial_k n^b.
\end{equation}
Substituting Eq.(\ref{10}) into Eq.(\ref{12}), one can obtain
\begin{equation}
H=\frac{1}{2\pi}\sum_{l=1}^{N}W_l \int_{L_l}A_i dx^i\label{helint1}.
\end{equation}
It can be seen that when these $N$ vortex filaments are $N$ closed
curves, i.e., a family of $N$ knots $\xi_l
(l=1,2,\cdot\cdot\cdot,N)$, Eq.(\ref{helint1}) leads to
\begin{equation}
H=\frac{1}{2\pi}\sum_{l=1}^{N}W_l \oint_{\xi_l}A_i dx^i.\label{15}
\end{equation}
This is a very important expression. Consider a transformation of
complex function $Z^{'}=e^{i\theta}Z$, this gives the U(1) gauge
transformation of $A_i: A_i^{'} = A_i + \partial_i\theta $, where
$\theta\in R$ is a phase factor denoting the U(1) gauge
transformation. It is seen that the $\partial_i\theta$ term in
Eq.(\ref{15}) contributes nothing to the integral $H$ when the
vortex filaments are closed, hence the expression (\ref{15}) is
invariant under the U(1) gauge transformation. As pointed out in
Ref.\cite{topo2} , a singular vortex filament is either closed ring
or infinite curve, therefore we conclude that the Hopf invariant is
a spontaneous topological invariant for the vortex filaments in
excitable media.

It is well known that many important topological numbers are related
to a knot family such as the self-linking number and Gauss linking
number. In order to discuss these topological numbers of knotted
vortex filaments, we define Gauss mapping:
\begin{equation}
\vec{m}: S^1 \times S^1 \rightarrow S^2,
\end{equation}
where $\vec{m}$ is a unit vector
\begin{equation}
\vec{m}(\vec{x},
\vec{y})=\frac{\vec{y}-\vec{x}}{|\vec{y}-\vec{x}|},\label{unit}
\end{equation}
where $\vec{x}$ and $\vec{y}$ are two points, respectively, on the
knots $\xi_k$ and $\xi_l$ (in particular, when $\vec{x}$ and
$\vec{y}$ are the same point on the same knot $\xi$ , $\vec{n}$ is
just the unit tangent vector $\vec{T}$ of $\xi$ at $\vec{x}$ ).
Therefore, when $\vec{x}$ and $\vec{y}$ , respectively, cover the
closed curves $\xi_k$ and $\xi_l$ once, $\vec{n}$ becomes the
section of sphere bundle $S^2$. So, on this $S^2$ we can define the
two-dimensional unit vector $\vec{e}=\vec{e}(\vec{x}, \vec{y})$.
$\vec{e}$, $\vec{m}$ are normal to each other, i.e. ,
\begin{eqnarray}
&&\vec{e}_1\cdot\vec{e}_2=\vec{e}_1\cdot\vec{m}=\vec{e}_2\cdot\vec{m}=0,
\nonumber\\&&\vec{e}_1\cdot\vec{e}_1=\vec{e}_2\cdot\vec{e}_2=\vec{m}\cdot\vec{m}=1.
\end{eqnarray}
In fact, the gauge field $\vec{A}$ can be decomposed in terms of
this two-dimensional unit vector $\vec{e}$:
$A_i=\epsilon_{ab}e^a\partial_i e^b-\partial_i\theta$, where
$\theta$ is a phase factor\cite{topo-cur1,topo-cur2}. Because the
$(\partial_i\theta)$ term does not contribute to the integral $H$,
$A_i$ can in fact be expressed as
\begin{equation}
A_i=\epsilon_{ab}e^a\partial_ie^b.
\end{equation}
Substituting it into Eq.(14), one can obtain
\begin{equation}
H=\frac{1}{2\pi}\sum_{k=1}^{N}W_k
\oint_{\xi_k}\epsilon_{ab}e^a(\vec{x},
\vec{y})\partial_ie^b(\vec{x}, \vec{y})dx^i.\label{hel}
\end{equation}
Noticing the symmetry between the points $\vec{x}$ and $\vec{y}$ in
Eq.(\ref{unit}), Eq.(\ref{hel}) should be reexpressed as
\begin{equation}
H=\frac{1}{2\pi}\sum_{k, l=1}^N W_k W_l \oint_{\xi_k}
\oint_{\xi_l}\epsilon_{ab}\partial_i e^a\partial_j e^bdx^i\wedge
dy^j.\label{hel2}
\end{equation}
In this expression there are three cases: (1) $\xi_k$ and $\xi_l$
are two different vortex filaments $(\xi_k\neq\xi_l)$, and $\vec{x}$
and $\vec{y}$ are therefore two different points
$(\vec{x}\neq\vec{y})$; (2) $\xi_k$ and $\xi_l$ are the same vortex
filaments $(\xi_k=\xi_l)$, but $\vec{x}$ and $\vec{y}$ are two
different points $(\vec{x}\neq\vec{y})$; (3) $\xi_k$ and $\xi_l$ are
the same vortex filaments $(\xi_k=\xi_l)$, and $\vec{x}$ and
$\vec{y}$ are the same points $(\vec{x}=\vec{y})$. Thus,
Eq.(\ref{hel2}) can be written as three terms:
\begin{eqnarray}
&&H=\sum_{k=1(k=l, \vec{x\neq}\vec{y})}^N \frac{1}{2\pi}W_k^2
\oint_{\xi_k} \oint_{\xi_k} \epsilon_{ab} \partial_i e^a\partial_j
e^b dx^i \wedge dy^j \nonumber\\&&+\frac{1}{2\pi}\sum_{k=1}^N W_k^2
\oint_{\xi_k} \epsilon_{ab} e^a\partial_i e^bdx^i
\nonumber\\&&+\sum_{k, l=1(k\neq l)}^N \frac{1}{2\pi}W_k W_l
\oint_{\xi_k} \oint_{\xi_l} \epsilon_{ab}
\partial_i e^a\partial_j e^b
dx^i \wedge dy^j.\label{hel3}
\end{eqnarray}
By making use of the relation
$\epsilon_{ab}\partial_ie^a\partial_je^b=\frac{1}{2}\vec{m}\cdot(\partial_i\vec{m}\times\partial_j\vec{m})$,
the Eq.(\ref{hel3}) is just
\begin{eqnarray}
H&=&\sum_{k=1(\vec{x}\neq\vec{y})}^N \frac{1}{4\pi}W_k^2
\oint_{\xi_k}\oint_{\xi_k} \vec{m}^*(dS) \nonumber \\ &&
+\frac{1}{2\pi}\sum_{k=1}^N W_k^2 \oint_{\xi_k} \epsilon_{ab}e^a
\partial_ie^bdx^i \nonumber \\ &&
+\sum_{k, l=1(k\neq l)}^N \frac{1}{4\pi} W_k W_l
\oint_{\xi_k}\oint_{\xi_l} \vec{m}^*(dS),\label{hel4}
\end{eqnarray}
where
$\vec{m}^*(dS)=\vec{m}\cdot(\partial_i\vec{m}\times\partial_j\vec{m})dx^i\wedge
dy^j(\vec{x}\neq\vec{y})$ denotes the pullback of the $S^2$ surface
element.

In the following we will investigate the three terms in the
Eq.(\ref{hel4}) in detail. Firstly, the first term of
Eq.(\ref{hel4}) is just related to the writhing number $Wr(\xi_k)$
of $\xi_k$\cite{wri,wri1}
\begin{equation}
Wr(\xi_k)=\frac{1}{4\pi}\oint_{\xi_k}\oint_{\xi_l}
\vec{m}^*(dS).\label{writhing}
\end{equation}
For the second term, one can prove that it is related to the
twisting number $Tw(\xi_k)$ of $\xi_k$
\begin{eqnarray}
\frac{1}{2\pi}\oint_{\xi_k}\epsilon_{ab}e^a\partial_ie^bdx^i
&&=\frac{1}{2\pi}\oint_{\xi_k}(\vec{T}\times\vec{V})\cdot
d\vec{V}\nonumber\\&&=Tw(\xi_k),\label{twisting}
\end{eqnarray}
where $\vec{T}$ is the unit tangent vector of knot $\xi_k$ at
$\vec{x}$ ($\vec{m}=\vec{T}$ when $\vec{x}=\vec{y}$) and $\vec{V}$
is defined as
$e^a=\epsilon^{ab}V^b(\vec{V}\perp\vec{T},\vec{e}=\vec{T}\times\vec{V})$.
In terms of the White formula\cite{wri,wri1}
\begin{equation}
SL(\xi_k)=Wr(\xi_k)+Tw(\xi_k),\label{self}
\end{equation}
we see that the first and the second terms of Eq.(\ref{hel4}) just
compose the self-linking numbers of knots.

Secondly, for the third term, one can prove that
\begin{eqnarray}
&&\frac{1}{4\pi}\oint_{\xi_k}\oint_{\xi_l}
\vec{m}^*(dS)\nonumber\\&&=\frac{1}{4\pi}\epsilon^{ijk}\oint_{\xi_k}dx^i\oint_{\xi_l}dy^j
\frac{(x^k-y^k)}{\|\vec{x}-\vec{y}\|^3}\nonumber\\&&=Lk(\xi_k,\xi_l)~~(k\neq
l),\label{linking}
\end{eqnarray}
where $Lk(\xi_k,\xi_l)$ is the Gauss linking number between $\xi_k$
and $\xi_l$\cite{gau}. Therefore, from Eqs.(\ref{writhing}),
(\ref{twisting}), (\ref{self}) and (\ref{linking}), we obtain the
important result:
\begin{equation}
H=\sum_{k=1}^N W_k^2 SL(\xi_k)+\sum_{k, l=1(k\neq
l)}^NW_kW_lLk(\xi_k, \xi_l).\label{helr}
\end{equation}
This precise expression just reveals the relationship between $H$
and the self-linking and the linking numbers of the vortex filaments
knots family. Since the self-linking and the linking numbers are
both the invariant characteristic numbers of the vortex filaments
knots family in topology, $H$ is an important topological invariant
required to describe the linked vortex filaments in excitable media.

So far, we obtained a more essential topological formulary of charge
density of knotted vortex filaments and revealed the topological
inner relationship of between the Hopf invariant and the
self-linking and the linking numbers of knotted vortex filaments. We
conclude that the Hopf invariant is just the sum of the linking and
self-linking numbers of the knotted vortex filaments. In the present
study, it should be pointed out that, when we discussed the
topological properties of the vortex filaments, the regular
condition $D^i(\frac{\phi}{x})\neq0$ must be satisfied. Now the
question is coming, when this condition fails, what will happen
about the vortex filaments? The answer is related to the evolution
of the vortex filaments\cite{topo-cur1,topo-cur2}. As we all known
that the evolution of the vortex filaments which include generating,
annihilating, splitting, or merging may obey some topological
constraints. These constraints usually relate to the topological
numbers of the vortex filaments, such as the topological charges and
linking numbers. So it is naturally to think that the precise
expression Eq.(\ref{helr}) of Hopf invariant may imply a new
topological constraint on the behavior of the knotted vortex
filaments. What the rigorous description of this possible new
constraint is will be investigated in our further works.

\end{document}